# Unitarity Constraint On Kinematical Analyses of the GSI Time-Modulated Radioactive Decay Experiment

Murray Peshkin*
*Physics Division, Argonne National Laboratory, Argonne, IL 60439*

Abstract

It is tempting to try to explain the reported time-modulated decay rate in the GSI experiment by arguing that the matrix element for decay at any time results in an interference between two slightly different momentum values in the parent ion beam. I show here that a unitarity theorem of quantum mechanics rules out a wide class of such explanations.

## The Unitarity Constraint

In the GSI experiment [1], hydrogen-like $^{140}$Pr atoms are reported to decay into the K-capture channel at a rate which is modulated at some frequency $\Omega$. The final state consists of a neutrino whose momentum wave function has two components because it is produced as an electron neutrino, which mixes two mass eigenstates. Let us assume that the matrix element for production of those two momentum states depends upon interference between two momentum values in the wave function of the Pr nucleus that differ by some value of **q** determined by the momentum difference between the two components of the neutrino wave function. It would then seem that there is an opportunity to have temporal oscillations in the decay rate with $\Omega$ determined by the neutrino mass difference. Theories based on that idea have been put forth [2,3]. Here I point out that a wide class of such theories cannot produce the wanted temporal oscillations.

The excluded theories are those in which the decay rate depends upon quantities $F(\mathbf{q},t)$ defined by

$$F(\mathbf{q},t) = \int \psi(\mathbf{p}+\mathbf{q},t)^* \psi(\mathbf{p},t) \, d^3\mathbf{p}, \qquad (1)$$

where $\psi(\mathbf{p},t)$ is the momentum-space wave function of the $^{140}$Pr ion, some wave packet originating at the time and place of the production of the ion, and **q** is a momentum shift determined by the particulars of the decay process. The decay rate may depend upon such $F(\mathbf{q},t)$ for a range of values of **q**. It has been suggested [1,2,3] that the kinematics of the process are such that the observed modulation frequency $\Omega$ is consistent with the kinematics of the decay process.

However, the unitarity principle of quantum mechanics implies that $F(\mathbf{q},t)$ is independent of the time $t$ because

$$\psi(\mathbf{p},t) = U(t)\psi(\mathbf{p},0) \qquad (2)$$

and similarly for $\psi(\mathbf{p}+\mathbf{q},t)$ with the same unitary operator $U(t)$. In other words, $F(\mathbf{q},t)$ is a constant of the motion for each $\mathbf{q}$, and there can be no temporal oscillations. That result is unchanged if one starts with an impure state represented by a density matrix at time $t=0$. Published kinematical analyses of which I am aware appear to be excluded by this unitary constraint.

In stating that unitarity forces $F(\mathbf{q},t)$ to be independent of $t$, I implicitly ignored the radioactive decay of the parent ion, since $U(t)$ is unitary only if $\psi(\mathbf{p},t)$ includes all decay channels as well as the $^{140}$Pr channel. Thus unitarity does not directly address the experiment; it addresses theories in which the effect of the radioactivity on the kinematics of the parent beam is not taken into account. That effect causes each momentum component of $\psi$ to have a small range of energies determined by the half life of the ion, and the consequences of that may need further investigation.

Finally, I note that none of this applies to the solar neutrino oscillations, which are spatial, not temporal. However, there is one similarity. Spatial oscillations are possible in the solar case only because the neutrino detector is smaller than the oscillation length and does not in effect average over the entire beam. In a GSI-like experiment, Eq.(1) applies strictly only to an experiment that looks uniformly at the entire beam. A final state detector that selects only part of the wave packet by sampling only part of the space occupied by the beam could in principle lead to a signal not given by Eq.(1), but that seems unlikely to be a practical possibility.

## Acknowledgements

I thank Harry J. Lipkin and John P. Schiffer for instructive conversations about this problem. This work was supported by the U.S. Department of Energy, Office of Nuclear Physics, under Contract No. DE-AC02-06CH11357.

* email: peshkin@anl.gov